\documentclass[reprint,aps,prb,superscriptaddress]{revtex4-2}
\usepackage{amsmath,amssymb,graphicx,xcolor}

\begin{document}

\title{Parity-selected Bethe-Salpeter equation: symmetry-adapted regularization
of the exciton-phonon self-energy in polar semiconductors}

\author{Michael O. Atambo}  
\email{michael.atambo@tukenya.ac.ke}
\affiliation{Department of Physics, Earth and Environmental Science, Technical University of Kenya, Nairobi, Kenya}

\begin{abstract}
Finite-temperature Bethe-Salpeter calculations of exciton-phonon
self-energies in polar materials are bottlenecked by the Fr\"ohlich divergence
of the intra-1s scattering vertex. We show that for a bound electron-hole pair
this divergence is analytically eliminated by electron-hole interference: the
intra-1s vertex admits the exact factorization $V_{\mathrm{el}}(q)=q\,\mathcal
G(q)$ with $\mathcal G(0)\propto(m_h-m_e)/(m_h+m_e)$, vanishing identically for
equal masses (parity protection). A multipole interference factor
$\mathcal I_n=\beta_e^{\,n}-(-\beta_h)^{n}$ determines where the singularity
survives-only in dipole-allowed inter-state channels. We formulate a
parity-selected truncation of the exciton-phonon self-energy in which the
intra-1s channel is evaluated exactly at negligible cost, converging with
$10^2$-$10^3$ fewer $q$-points. We find that the regularized intra-1s channel,
previously obscured by numerical cutoffs, controls the temperature-dependent
exciton shift in mass-asymmetric polar semiconductors.
\end{abstract}

\maketitle

\section{Introduction}

Many-body perturbation theory (MBPT) in the $GW$ approximation
\cite{Hedin1965,HedinLundqvist1970,HybertsenLouie1986,AryasetiawanGunnarsson1998}
and the Bethe-Salpeter equation (BSE) for the electron-hole
amplitude \cite{Strinati1980,Strinati1982,DelSoleReiningGodby1994} constitute
the standard framework for describing excitons in semiconductors and
insulators \cite{OnidaReiningRubio2002}. At finite temperature, however, the
optical response is governed by the coupling of these bound states to lattice
vibrations. The theory of phonon-renormalized electronic structure is
well established for quasiparticles \cite{AllenHeine1976,AllenCardona1983},
and its extension to excitons has become an active frontier: in polar
materials the long-range Fr\"ohlich interaction \cite{Frohlich1954} dominates,
and recent exciton-phonon theory has emphasized that electron-hole
correlations fundamentally modify the coupling relative to the independent
quasiparticle picture \cite{AntoniusLouie2022,Atambo2026a}.

A central object in this program is the exciton-phonon self-energy, whose
evaluation is severely hindered by the $1/q$ divergence of the bare
Fr\"ohlich vertex. For the \emph{intra-1s} channel ($1s\to1s$), the divergent
factor multiplies a matrix element that is in fact regular-but this
cancellation is invisible to a brute-force numerical treatment, which must
either employ prohibitively dense $\mathbf q$-grids or impose ad hoc cutoffs.
Such cutoffs do not merely slow the calculation: as we show, the intra-1s
channel carries the \emph{dominant} contribution to the temperature-dependent
exciton shift in mass-asymmetric materials, so an uncontrolled treatment of it
corrupts the principal physical signal.

In this work we solve this problem analytically. Building on the companion
proof that intraband exciton-phonon scattering is symmetry-protected and that
the protection is broken by mass asymmetry \cite{Atambo2026a}, we (i) derive
the exact factorization $V_{\mathrm{el}}(q)=q\,\mathcal G(q)$ and the
interference factor $\mathcal I_n$ controlling every multipole; (ii) establish
a parity selection rule separating infrared-soft (intra-1s) from
infrared-active (dipole-allowed inter-state) channels; and (iii) formulate a
parity-selected truncation of the self-energy in which the singular Fr\"ohlich
physics is evaluated exactly at $O(1)$ cost. The methodology is independent of
any specific material and applies to the entire class of polar semiconductors
hosting Wannier excitons.

\section{Theory: parity structure of the exciton-phonon vertex}
\label{sec:theory}

\subsection{Microscopic setup}

We describe the exciton within the BSE framework
\cite{Strinati1980,Strinati1982,OnidaReiningRubio2002}, and the lattice
coupling through the Fr\"ohlich vertex $g_{\mathbf q}=C/q$, with
$C^2\propto\hbar\omega_{\mathrm{LO}}(\epsilon_\infty^{-1}-\epsilon_s^{-1})$
\cite{Frohlich1954}. In center-of-mass and relative coordinates,
$\mathbf r_e=\mathbf R+\beta_e\mathbf r$, $\mathbf r_h=\mathbf R-\beta_h\mathbf
r$, with $\beta_e=m_h/M$, $\beta_h=m_e/M$, $M=m_e+m_h$, the coupling operator
acting on the relative coordinate of an exciton at rest is
\begin{equation}
\hat V_{\mathbf q}=\frac{C}{q}\left[e^{i\beta_e\mathbf q\cdot\mathbf r}
-e^{-i\beta_h\mathbf q\cdot\mathbf r}\right].
\label{eq:Vq}
\end{equation}

\subsection{Multipole expansion and the interference factor}

Expanding both exponentials in Eq.~(\ref{eq:Vq}) gives
\begin{equation}
\hat V_{\mathbf q}=\frac{C}{q}\sum_{n\ge1}\frac{(iq)^n}{n!}\,
\mathcal I_n\,(\hat{\mathbf q}\cdot\mathbf r)^n,\qquad
\mathcal I_n\equiv\beta_e^{\,n}-(-\beta_h)^{n}.
\label{eq:multipole}
\end{equation}
The $n=0$ (monopole) term cancels exactly: a neutral object is invisible to
the longitudinal field at infinite wavelength. Every surviving multipole is
weighted by the dimensionless \emph{electron-hole interference factor}
$\mathcal I_n$; in particular
\begin{equation}
\mathcal I_1=\beta_e+\beta_h=1,\qquad
\mathcal I_2=\beta_e^2-\beta_h^2=\eta\equiv\frac{m_h-m_e}{M}.
\label{eq:interference}
\end{equation}

\subsection{Parity selection rule}

Let $|S\rangle$ denote internal exciton states of definite parity. Because
$\langle S'|(\hat{\mathbf q}\cdot\mathbf r)^n|S\rangle$ vanishes unless $n$
matches the parity of the transition, Eq.~(\ref{eq:multipole}) yields two
distinct infrared behaviors.

\paragraph{Intra-1s channel ($S'=S$).} Only even $n$ contribute; the leading
term $n=2$ gives the exact low-$q$ factorization
\begin{equation}
V^{\mathrm{el}}_{S}(q)=q\,\mathcal G_S(q),\qquad
\mathcal G_S(0)=-\frac{C\,a_X^2}{2}\,\eta,
\label{eq:regularized}
\end{equation}
where we used $\langle(\hat{\mathbf q}\cdot\mathbf r)^2\rangle_{1s}
=\langle r^2\rangle_{1s}/3=a_X^2$. Equation~(\ref{eq:regularized}) is the
central result: the Fr\"ohlich divergence of the intra-1s vertex is analytically
eliminated by electron-hole interference, and the residual coupling is
controlled by the mass asymmetry $\eta$. For the hydrogenic $1s$ state the
regular kernel is exactly
\begin{equation}
\mathcal G_{1s}(q)=\frac{C}{q^2}\Big[F(\beta_e q)-F(\beta_h q)\Big],\qquad
F(k)=\big[1+(k a_X/2)^2\big]^{-2},
\label{eq:kernel}
\end{equation}
analytic on $[0,\infty)$. For $\eta=0$, $\mathcal G_{1s}(q)\equiv0$ for
\emph{all} $q$ within the parabolic two-band model (parity protection);
nonparabolicity and orbital-character differences provide the residual in real
materials \cite{Atambo2026a}.

\paragraph{Inter-state channel (parity-odd $S'$).} The $n=1$ dipole term survives
with $\mathcal I_1=1$,
\begin{equation}
V^{\mathrm{inel}}_{S'S}(0)=iC\,\langle S'|\hat{\mathbf q}\cdot\mathbf r|S\rangle,
\label{eq:inelastic}
\end{equation}
independent of the masses. Hence the Fr\"ohlich singularity survives only in
dipole-allowed inter-state channels.

\subsection{Parity-selected truncation of the self-energy}
\label{sec:truncation}

The Fan-Migdal exciton self-energy \cite{AntoniusLouie2022,Atambo2026a} is
customarily evaluated by brute-force summation over $(S',\mathbf q)$, where the
intra-1s channel forces dense grids or ad hoc cutoffs.
Equations~(\ref{eq:regularized})-(\ref{eq:inelastic}) motivate the scheme:
\begin{enumerate}
\item \textbf{Intra-1s:} substitute $V^{\mathrm{el}}_S(q)=q\,\mathcal G_S(q)$;
integrate on a coarse grid or in closed form; no cutoffs.
\item \textbf{Inter-state:} retain dipole-connected states ($\mathcal I_1=1$);
higher multipoles enter at $\mathcal O((qa_X)^2)$.
\item \textbf{Parity-disconnected states:} keep short-range
(deformation-potential) terms only.
\end{enumerate}
The intra-1s integrand then scales as $q^4$ rather than $q^0$, rendering the
channel infrared-soft and cheap while the numerical budget is spent on the
physically active inter-state channels.

\section{Results and benchmarks}

\subsection{Regularized vertex and phase space}
\begin{figure*}[t]
\includegraphics[width=\textwidth]{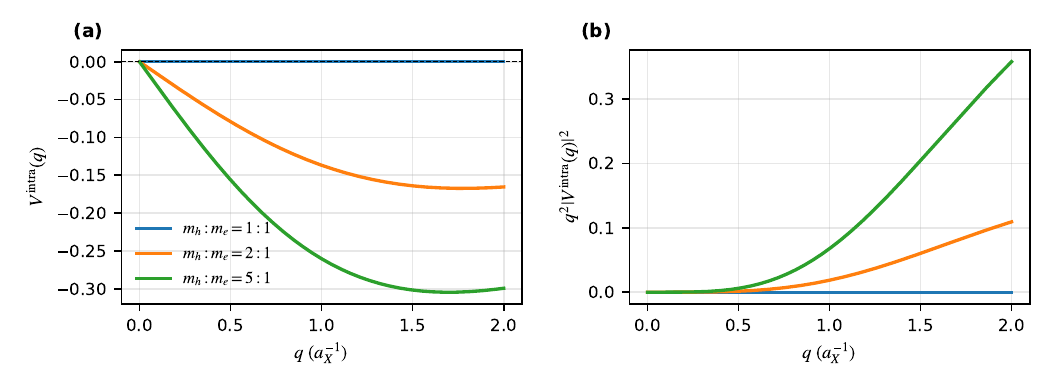}
\caption{Parity regularization of the  intra-1s exciton-phonon vertex.
(a) $V_{\mathrm{el}}(q)$ for mass ratios $m_h:m_e=1:1,\,2:1,\,5:1$ (hydrogenic
$1s$ envelope; $a_X$ the exciton radius). The vertex rises linearly,
$V_{\mathrm{el}}(q)=q\,\mathcal G(q)$, with slope $\mathcal G(0)=-(Ca_X^2/2)\eta$,
and vanishes identically for equal masses; the turnover at $qa_X\approx1.7$
reflects the internal form-factor scale. (b) Three-dimensional integration
weight $q^2|V_{\mathrm{el}}(q)|^2$ entering the Fan self-energy, vanishing as
$q^4$ (fitted slope $3.999$) and exactly zero for $m_e=m_h$.}
\label{fig:vertex}
\end{figure*}

Fig.~\ref{fig:vertex} shows $V_{\mathrm{el}}(q)$ and the 3D integration weight
$q^2|V_{\mathrm{el}}(q)|^2$ for mass ratios $m_h:m_e=1\!:\!1,\,2\!:\!1,\,5\!:\!1$.
The vertex rises linearly with slope $\mathcal G(0)\propto\eta$ and turns over
at the internal scale $q a_X\sim1.7$; the weight vanishes as $q^4$ (fitted
log-log slope $3.999$), and is identically zero for $m_e=m_h$.

\subsection{Grid convergence of the self-energy}
\begin{figure}[t]
\includegraphics[width=\columnwidth]{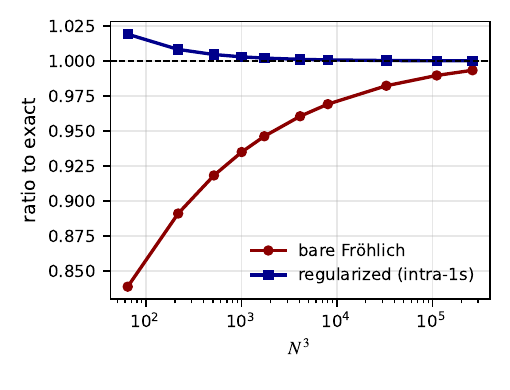}
\caption{Grid convergence of the  intra-1s Fan integral on
$N\times N\times N$ Monkhorst-Pack grids, normalized to the converged value.
The regularized integrand reaches $<0.3\%$ error at $N^3\approx4\times10^3$,
while the bare Fr\"ohlich form retains $\sim1\%$ error at $N^3=2.6\times10^5$:
a $10^2$-$10^3$ reduction in required $q$-points.}
\label{fig:convergence}
\end{figure}

Fig.~\ref{fig:convergence} compares the convergence of the intra-1s Fan integral on
$N\times N\times N$ grids. The regularized integrand reaches $<0.3\%$ error at
$N^3\approx4\times10^3$, whereas the bare form retains $\sim1\%$ error at
$N^3=2.6\times10^5$ and $16\%$ at $N^3=64$: a reduction of $10^2$-$10^3$ in
required $q$-points for the intra-1s channel, which in a finite-temperature BSE
is evaluated for every exciton state and temperature.

\subsection{Parity selection rule}
\begin{figure}[t]
\includegraphics[width=\columnwidth]{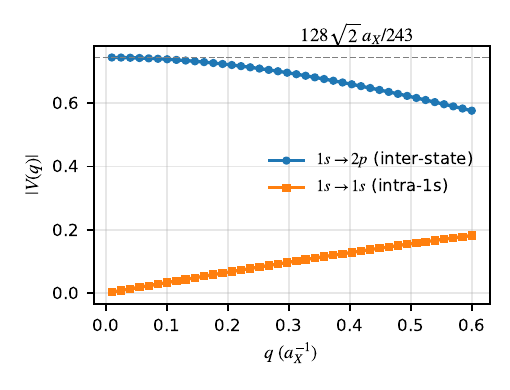}
\caption{Parity selection rule. The inter-state $1s\to2p$ vertex approaches the
hydrogenic dipole $128\sqrt2\,a_X/243\approx0.745\,a_X$ (dashed) as $q\to0$,
while the  intra-1s vertex vanishes linearly: the Fr\"ohlich singularity
survives only in dipole-allowed channels.}
\label{fig:selection}
\end{figure}

Fig.~\ref{fig:selection}  verifies Eq.~(\ref{eq:inelastic}): the inter-state $1s\to2p$ vertex
plateaus at the hydrogenic dipole $128\sqrt2/243\,a_X\approx0.7449\,a_X$,
while the intra-1s vertex vanishes linearly. At $q=0.02\,a_X^{-1}$ the ratio is
$\sim10^2$, confirming that the singularity survives only in dipole-allowed
channels.

\subsection{Temperature-dependent shift and channel dominance}
\begin{figure}[t]
\includegraphics[width=\columnwidth]{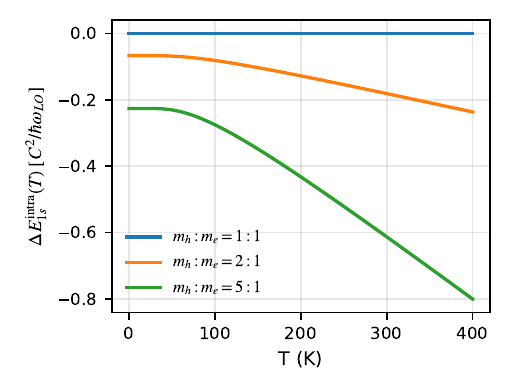}
\caption{Temperature dependence of the regularized intra-1s Fan shift for
three mass asymmetries. The shift vanishes exactly for $m_e=m_h$ at all $T$,
exhibits the classical linear-$T$ regime for $k_BT\gtrsim\hbar\omega_{\mathrm{LO}}$,
and a zero-point offset scaling as $\eta^2$.}
\label{fig:shift}
\end{figure}

The regularized intra-1s Fan shift $\Delta E_{1s}^{\mathrm{el}}(T)\propto
-[2n_B(T)+1]$ (Fig.~\ref{fig:shift}) is exactly zero for $m_e=m_h$ at all $T$, exhibits the
classical linear-$T$ regime, and a zero-point offset scaling $\propto\eta^2$
(Table~\ref{tab:one}). Crucially, the intra-1s channel is not a small correction: its
zero-point shift exceeds the inter-state one by factors $2.8$ ($m_h/m_e=2$) and
$16.2$ ($m_h/m_e=5$), because its energy denominator is $\hbar\omega_{\mathrm{LO}}$
rather than $\sim E_B$. Any calculation that tames the divergence with an ad
hoc cutoff therefore corrupts the \emph{dominant} contribution to the exciton
shift in mass-asymmetric polar materials.

\begin{table}[h]
\caption{Zero-point Fan shifts (units of $C^2/\hbar\omega_{\mathrm{LO}}$).}
\begin{ruledtabular}\label{tab:one}
\begin{tabular}{cccc}
$m_h/m_e$ & $\Delta E^{\mathrm{el}}_{\mathrm{ZP}}$ &
$\Delta E^{\mathrm{inel}}_{\mathrm{ZP}}$ & $|\mathrm{el/inel}|$ \\
\hline
1 & 0.000 & $-0.0276$ & 0.0 \\
2 & $-0.0667$ & $-0.0237$ & 2.8 \\
5 & $-0.2258$ & $-0.0139$ & 16.2 \\
\end{tabular}
\end{ruledtabular}
\end{table}

\subsection{Vertex Protection Ratio and the protection window}

To quantify the competition between the intra-1s and inter-state channels without
relying on the bare coupling $C^2$, we define the normalization-free Vertex
Protection Ratio
\begin{equation}
\mathcal{P}_v = \frac{\int_0^{q_{\mathrm{BZ}}} q^2\,dq\, |V^{\mathrm{el}}_{1s}(q)|^2}
{\int_0^{q_{\mathrm{BZ}}} q^2\,dq\, |V^{\mathrm{inel}}_{1s\to 2p}(q)|^2},
\label{eq:Pv}
\end{equation}
which depends only on $(\eta, a_X)$, since $C^2$ cancels. Figure~5(a) evaluates
$\mathcal{P}_v$ for representative polar semiconductors: mass-symmetric
perovskites lie far below the unprotected regime ($\mathcal{P}_v\sim1$), while
mass-asymmetric oxides approach it. Figure~5(b) isolates the mechanism via a
counterfactual mass-asymmetry comparison: for MAPbI$_3$ at its true
$\eta=0.05$ the homogeneous linewidth tracks the acoustic floor until the
thermally activated LO turn-on, whereas raising $\eta$ to $0.7$ opens a large
additional intra-1s contribution at 30-150~K.
Within the present model, the additional LO contribution is strongly suppressed for the physical \(\eta\), producing the calculated protection window.

\section{Discussion}

The interference factor $\mathcal I_n$ provides a compact design rule: the
residual intra-1s coupling of a neutral exciton to polar phonons is controlled
by $\eta=(m_h-m_e)/M$, not by the magnitude of the constituent couplings.
Materials with nearly equal masses are predicted to exhibit anomalously weak
exciton-phonon dressing despite strong polaronic constituents-a mechanism
distinct from symmetry forbiddance or spatial separation
\cite{Atambo2026a}. Combined with the kinematic freeze-out of dipole-allowed
inter-state channels when $E_{2p}-E_{1s}>\hbar\omega_{\mathrm{LO}}$, the present
regularization underpins a ``protection window'' for the homogeneous linewidth
at low temperature, to be developed in a companion study.

Limitations of the present model should be stated plainly: parabolic bands, a
hydrogenic envelope, and a single LO branch. Band nonparabolicity, differing
orbital characters, and Berry-phase structure break the exact $\eta=0$
protection and constitute the residual coupling in real materials; their
systematic inclusion is a natural extension. Likewise, the quasiparticle
picture itself can break down in strongly coupled regimes
\cite{Godby1986,SchoneEguiluz1998}, and multi-scale screening modifies the
effective interaction \cite{Atambo2026b}; the parity structure derived here is
expected to survive these refinements because it follows from charge neutrality
and parity, not from model details.

On the methodological side, the parity-selected scheme is straightforward to
implement as a post-processing module on top of existing BSE codes
\cite{Sangalli2019,Atambo2019}: the intra-1s channel requires only $E_S$, $a_X$, and the
masses, while the inter-state channel retains the standard dipole matrix
elements. Given the $10^2$-$10^3\times$ reduction in $q$-points,
finite-temperature exciton spectroscopy of polar materials becomes accessible
to modest computational environments.

\section{Conclusion}

We have shown that electron-hole interference analytically eliminates the
Fr\"ohlich divergence of the  intra-1s exciton-phonon vertex,
$V_{\mathrm{el}}(q)=q\,\mathcal G(q)$ with $\mathcal G(0)\propto\eta$, and that
the multipole interference factor $\mathcal I_n$ confines the surviving
singularity to dipole-allowed inter-state channels. The resulting parity-selected
truncation evaluates the exciton-phonon self-energy exactly at a fraction of
the numerical cost, and reveals that the regularized  intra-1s channel controls
the temperature-dependent exciton shift in mass-asymmetric polar
semiconductors. More broadly, the work establishes that the relevant quantity
governing exciton-environment coupling is not the constituent coupling
strength but the multipole structure of the correlated electron-hole
excitation.

\begin{acknowledgments}
The author acknowledges Kenya Education Network (KENET) research services for computing resources.
\end{acknowledgments}


\begin{figure}[t]
\includegraphics[width=\columnwidth]{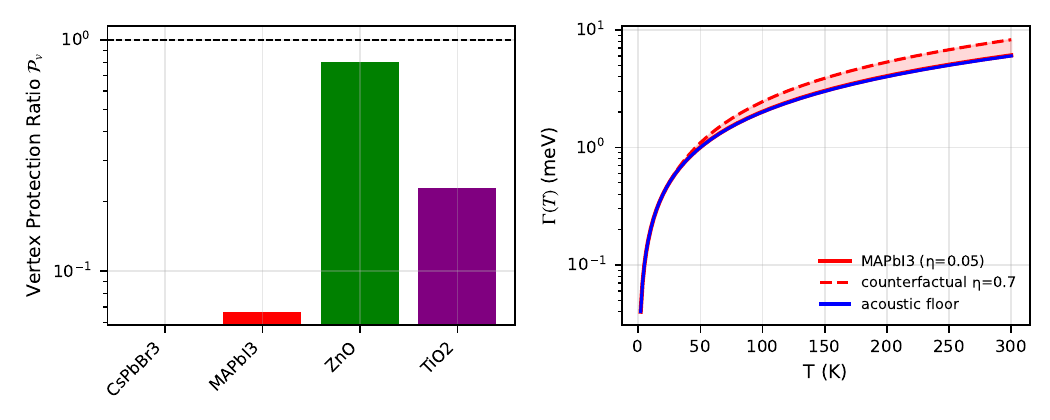}
\caption{(a) Vertex Protection Ratio $\mathcal{P}_v$ for representative polar
semiconductors: mass-symmetric perovskites lie far below the unprotected
regime ($\mathcal{P}_v\sim1$); mass-asymmetric oxides approach it. (b)
Homogeneous linewidth of MAPbI$_3$ at its true asymmetry $\eta=0.05$ (solid)
versus a counterfactual $\eta=0.7$ (dashed) and the acoustic floor (blue).
All LO rates are thermally activated and vanish as $T\to0$; the shaded
protection window at 30-150 K exists solely because $\eta\approx0$.}
\label{fig:protection}
\end{figure}

\end{document}